\documentclass[12pt]{article}
\usepackage{epsfig}
\usepackage{graphicx}  % standard LaTeX graphics tool

\newcommand{\simlt}
{\mbox{\raisebox{-0.5ex}{$\textstyle \; \sim$}
\raisebox{ 0.8ex}{$\textstyle  \!\!\!\!\!\!\! <$  }}}
\begin{document}
\title{Comparison of the Black Hole and Fermion Ball
Scenarios of the Galactic Center}
\author{Neven Bili\'{c}\thanks{Permanent
 address:
Rudjer Bo\v skovi\'c Institute,
P.O. Box 180, 10002 Zagreb, Croatia;
 \hspace*{5mm} Email: bilic@thphys.irb.hr}\ ,
Faustin  Munyaneza,
Gary B.\ Tupper, \\and Raoul D.\ Viollier\thanks{
 Email: viollier@physci.uct.ac.za}
\\
Institute of Theoretical Physics and Astrophysics, \\
 Department of Physics, University of Cape Town,  \\
 Private Bag, Rondebosch 7701, South Africa \\
 }
\maketitle

\begin{abstract}
After a discussion of the properties of degenerate fermion balls, we
analyze the orbit of the star S0-1, which has a
projected distance of $\sim$ 5 light-days to Sgr A$^{*}$, in the
supermassive black hole as well as
in the fermion ball scenarios of the Galactic center. It is shown that both
scenarios are consistent with the data, as measured during the last 6 years by
Genzel and coworkers and by Ghez and coworkers. The free parameters of the
projected orbit of a star are the unknown components of its velocity $v_{z}$
and distance $z$ to Sgr A$^{*}$ in 1995.4, with the $z$-axis being in the line
of sight. We show, in the case of S0-1, that the $z - v_{z}$
phase-space, which fits the data, is much larger for the fermion ball than for
the black hole scenario. Future measurements of the positions or radial
velocities of S0-1 and S0-2, which could be orbiting within such a fermion
ball, may reduce this allowed phase space and
eventually rule out one of the currently acceptable scenarios.
This could shed
some light on the nature of the supermassive compact dark object, or dark
matter in general, at the center of our Galaxy.
\end{abstract}

\section{Introduction}
Self-gravitating degenerate neutrino matter has been suggested as a model
for quasars, with neutrino masses in the 0.2 keV $\simlt m \simlt$ 0.5 MeV
range \cite{mark1} even before the black hole hypothesis of the quasars was
conceived \cite{lyn16}. More recently, supermassive compact dark objects consisting
of weakly interacting degenerate fermionic matter, with fermion masses in the
10 $\simlt m$/keV $\simlt$ 20 range, have been proposed
\cite{viol2,bil3,bil4,tsik5,mun6} as an alternative to the supermassive black
holes that are believed to reside at the centers of many galaxies.

The masses of $\sim$ 20 supermassive compact dark objects
at the
centers of inactive galaxies \cite{ho7} have been measured so far.
The most massive compact dark
object ever observed is located at the center of M87 in the Virgo cluster, and
it has a mass of $\sim$ 3 $\times$ 10$^{9} M_{\odot}$ \cite{macc8}.
NGC 3115, NGC 4594 and NGC 4374 are galaxies harbouring compact dark objects
with the next smaller mass of $\sim$ 10$^{9} M_{\odot}$.
If we
identify the object of maximal mass with a degenerate fermion ball at the
Oppenheimer-Volkoff (OV) limit \cite{opp9}, i.e. $M_{\rm OV}$ = 0.54$M_{\rm
Pl}^{3} \; m^{-2} g^{-1/2} \simeq$ 3 $\times$ 10$^{9} M_{\odot}$ \cite{bil4},
where $M_{\rm Pl} = \sqrt{\hbar c/G}$, this allows us to fix the fermion mass
to $\simeq$ 15 keV for a spin and particle-antiparticle degeneracy factor of
$g$ = 2. Such a relativistic object would have a radius of $R_{\rm OV}$ = 4.45
$R_{\rm S} \simeq$ 1.5 light days, where $R_{\rm S}$ is the Schwarzschild
radius of the mass $M_{\rm OV}$. It would thus be virtually indistinguishable
from a black hole of the same mass, as the closest stable orbit around a black
hole has a radius of 3 $R_{\rm S}$ anyway.

At the lower end of the observed mass range are the compact dark objects
located at the center of NGC 4945, M32 and our Galaxy \cite{eck10} with
masses of about 1,3 and 2.6
million solar masses, respectively. Interpreting the Galactic object as
a degenerate fermion
ball consisting of $m \simeq$ 15 keV and $g$ = 2 fermions, the radius is
$R_{\rm c} \simeq$ 21 light-days $\simeq$ 7 $\times$ 10$^{4} R_{\rm S}$
\cite{viol2}, $R_{\rm S}$ being the Schwarzschild radius of the mass
$M_{\rm c}$ = 2.6 $\times$ 10$^{6} M_{\odot}$. Such a nonrelativistic
object is far from being a black hole.
\begin{figure}[h]
\begin{center}
\includegraphics[width=.8\textwidth]{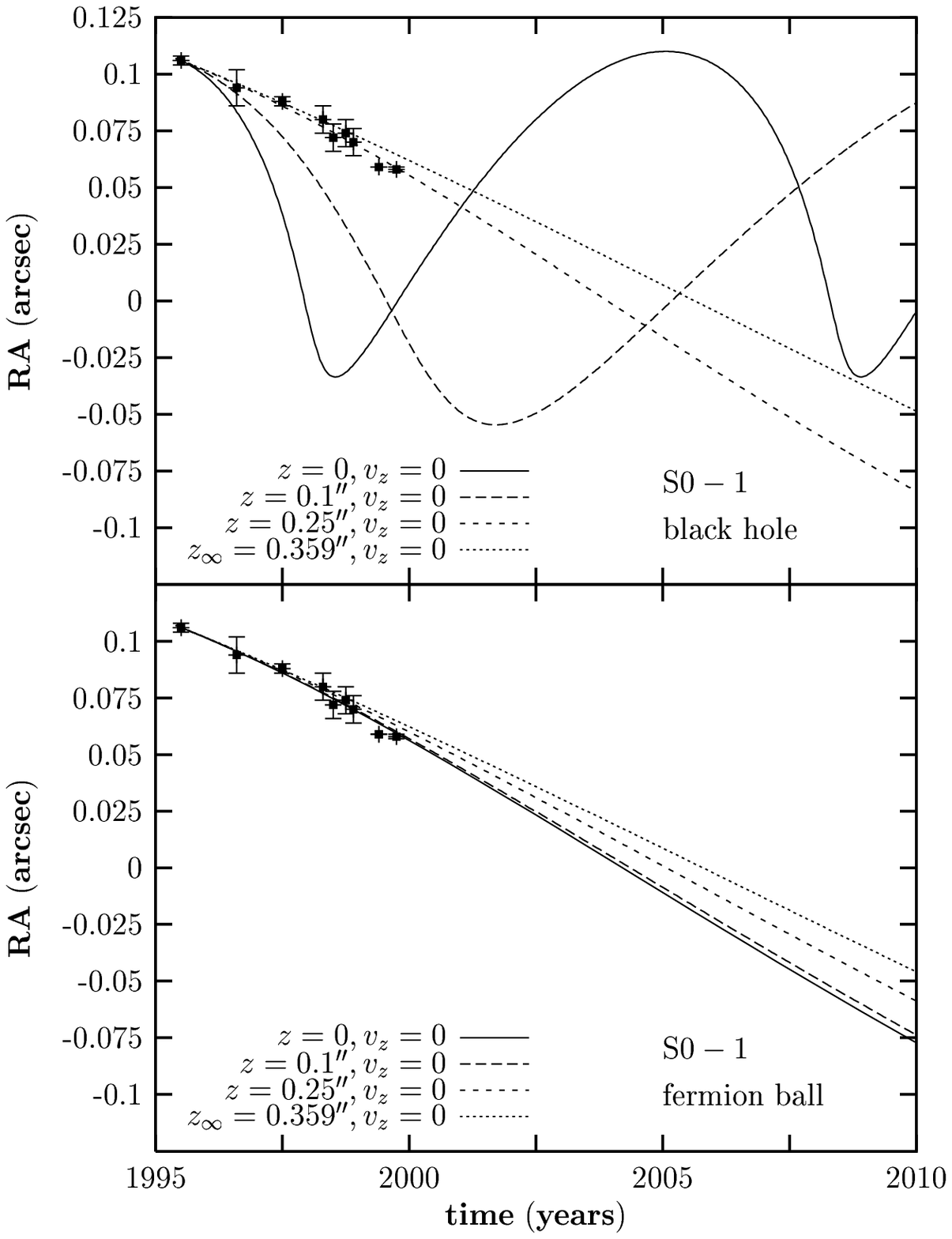}
\end{center}
\caption{
Right ascension of S0-1 versus time for various $|z|$ and $v_{x}$ =
340 km s$^{-1}$,
$v_{y}$ = - 1190 km s$^{-1}$ and $v_{z}$ = 0 in 1995.4.
}
\label{fig1a}
\end{figure}
The observed motion of stars within a projected distance of
$\sim$ 5 to $\sim$ 50 light-days from Sgr A$^{*}$ \cite{eck10},
the powerful and enigmatic
radio source at the Galactic center, yields, apart from the mass, an upper
limit for the radius of the fermion ball $R_{\rm c} \simlt$ 22 light
days. Matter orbiting in an optically thick and geometrically thin accretion
disk in or around such a fermion ball will only emit radiation at distances
larger than $\sim$ 10 mpc from the center, as both the density and the
circular frequency become nearly constant near the center of the
fermion ball \cite{bil3}.
The spectrum emitted by the disk will thus have a cut-off at frequencies
larger than $\sim$ 10$^{13}$ Hz, as is actually observed. Of course, there
will be a pile-up and instability of matter within $\sim$ 10 mpc, perhaps leading to the
formation of stars, as the gravitational tidal forces on nascent stars is much
smaller in the fermion ball than in the black hole scenario. These stars may be
eventually ejected from the central star cluster by intruder stars in close
binary encounters. The formation of such a fermion ball as well as its coexistence
at finite temperature with a Galactic halo composed of the same fermions
has been discussed by Lindebaum \cite{lind17} and Bili\'{c} \cite{bil18}
respectively, at this conference.
\begin{figure}[h]
\begin{center}
\includegraphics[width=.8\textwidth]{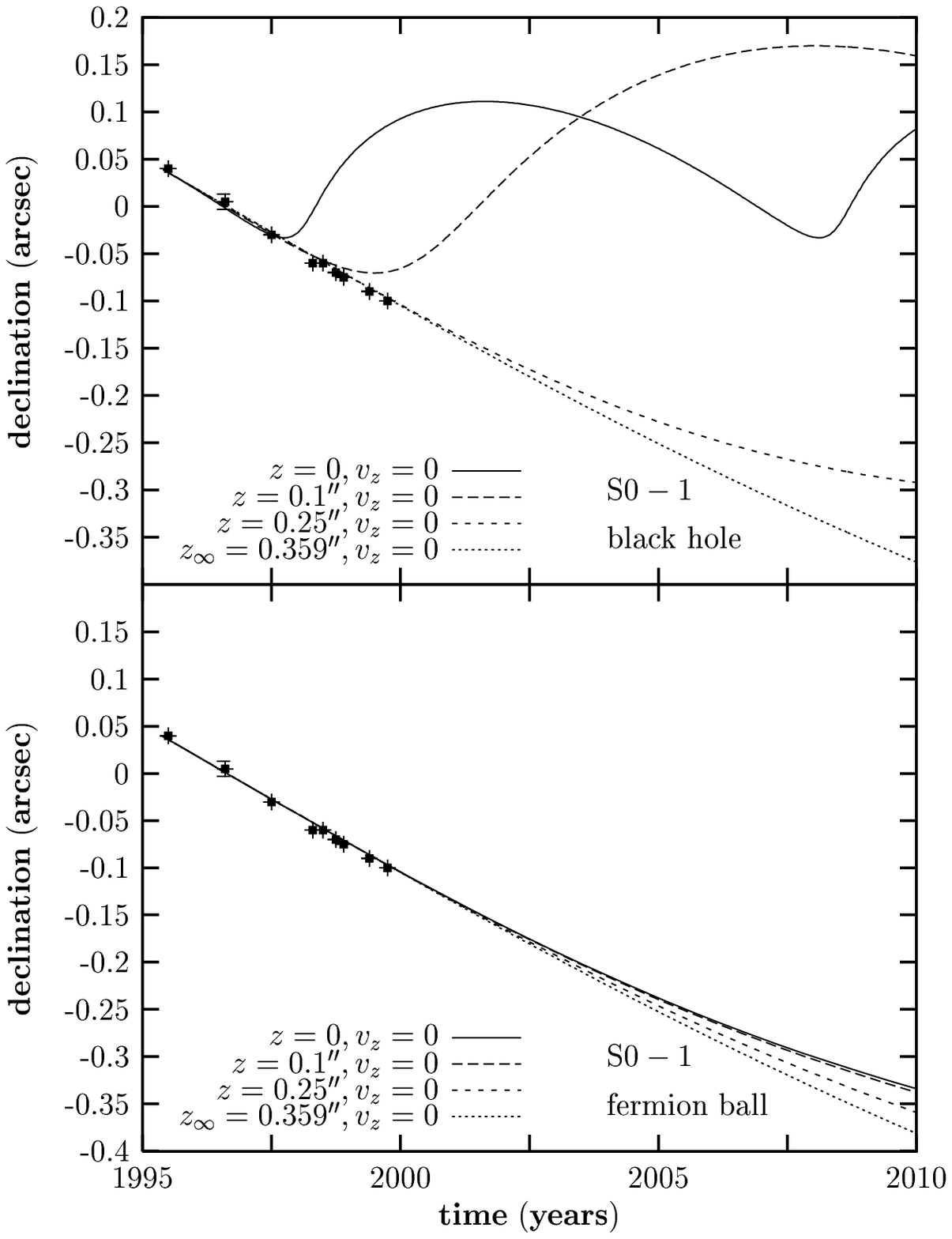}
\end{center}
\caption{
Declination  of S0-1 versus time for various $|z|$ and $v_{x}$ =
340 km s$^{-1}$,
$v_{y}$ = - 1190 km s$^{-1}$ and $v_{z}$ = 0 in 1995.4.
}
\label{fig2a}
\end{figure}
The required weakly interacting fermion of $\sim$ 15 keV mass
cannot be an active neutrino, as it would overclose the Universe by orders of
magnitude
\cite{kolb11}.
However, the $\sim$ 15 keV fermion could very well be a sterile neutrino,
contributing $\Omega_{\rm d} \simeq$ 0.3 to the dark matter fraction of the
critical density today. Indeed, as has been shown for an initial lepton
asymmetry of $\sim$ 10$^{-3}$, a sterile neutrino of mass $\sim$ 10 keV may be
resonantly produced in the early Universe with near closure density, i.e.
$\Omega_{\rm d} \sim$ 1
\cite{shi12}.
As an alternative possibility, the required $\sim$ 15 keV fermion could be the
axino
\cite{goto13}
or the gravitino
\cite{lyth14}
in soft supersymmetry breaking scenarios.

\section{Dynamics of the Stars Near the Galactic Center}
We now would like to compare the predictions of the black hole and fermion
ball scenarios of the Galactic center, for the stars with the smallest
projected distances to Sgr A$^{*}$, based on the measurements of their
positions during the last six years \cite{eck10}. The projected orbits of
three stars, S0-1 (S1), S0-2 (S2) and S0-4 (S8), show deviations from uniform
motion on a straight line during the last six years, and they thus may contain
nontrivial information about the potential. For our analysis we have selected
the star, S0-1, because its projected distance from Sgr A$^{*}$ in 1995.53,
4.4 mpc or 5.3 light-days, makes it most likely that it could be orbiting
within a fermion ball of radius $\sim$ 18 mpc or $\sim$ 21 light-days. We thus
may in principle distinguish between the black hole and fermion ball scenarios
for this star.

The dynamics of the stars in the gravitational field of the supermassive
compact dark object can be studied solving Newton's equation of motion, taking
into account the initial position and velocity vectors at, e.g., $t_{0}$ =
1995.4 yr, i.e., $\vec{r} (t_{0}) \equiv (x, y, z)$ and
$\dot{\vec{r}} (t_{0}) \equiv (v_{x}, v_{y}, v_{z})$. For the fermion ball the
source of gravitational field is the mass ${\cal{M}} (r)$ enclosed within a
radius $r$ \cite{bil4,mun6} while for the black hole it is $M_{\rm c} =
{\cal{M}}
(R_{\rm c})$ = 2.6 $\times$ 10$^{6} M_{\odot}$. The $x$-axis is chosen in the
direction opposite to the right ascension (RA), the $y$-axis in the direction
of the declination, and the $z$-axis points towards the sun. The black hole
and the center of the fermion ball are assumed to be at the position of Sgr
A$^{*}$ which is also the origin of the coordinate system at an assumed
distance of 8 kpc from the sun.

In Figs. 1 and 2 the right ascension (RA) and declination of S0-1 are plotted
as a function of time for various unobservable $z$'s and $v_{z}$ = 0 in
1995.4, in the black hole and fermion ball scenarios. The velocity components
$v_{x}$ = 340 km s$^{-1}$ and $v_{y}$ = - 1190 km s$^{-1}$ in 1995.4 have been
fixed from observations. In the case of a black hole, both RA and declination
depend strongly on $z$ in 1995.4, while the $z$-dependence of these quantities
in the fermion ball scenario is rather weak. We conclude that the RA and
declination data of S0-1 are well fitted with $|z| \approx$ 0.25'' in the
black hole scenario, and with $|z| \simlt$ 0.1'' in the fermion ball case
(1'' = 38.8 mpc = 46.2 light-days at 8 kpc). Of course, we can also try to fit
the data varying both the unknown radial velocity $v_{z}$ and the
unobservable radial distance $z$. The results are summarized in Fig. 3, where
the $z - v_{z}$ phase-space of 1995.4, that fits the data, is shown. The
small range of acceptable $|z|$ and $|v_{z}|$ values in the black hole
scenario (solid vertical line) reflects the fact that the orbit of S0-1 depend
strongly on $z$. The weak sensitivity of the orbit on $z$ in the fermion ball
case is the reason for the much larger $z - v_{z}$ phase-space fitting the
data of S0-1 \cite{eck10}, as shown by the dashed box. The dashed and solid
curves describe the just bound orbits in the fermion ball and black hole
scenarios,
respectively. The star S0-1 is unlikely to be unbound, because in the absence
of close encounters with stars of the central cluster, S0-1 would have to fall
in with an initial velocity that is inconsistent with the velocity dispersion
of the stars at infinity.
\begin{figure}[h]
\begin{center}
\includegraphics[width=0.9\textwidth]{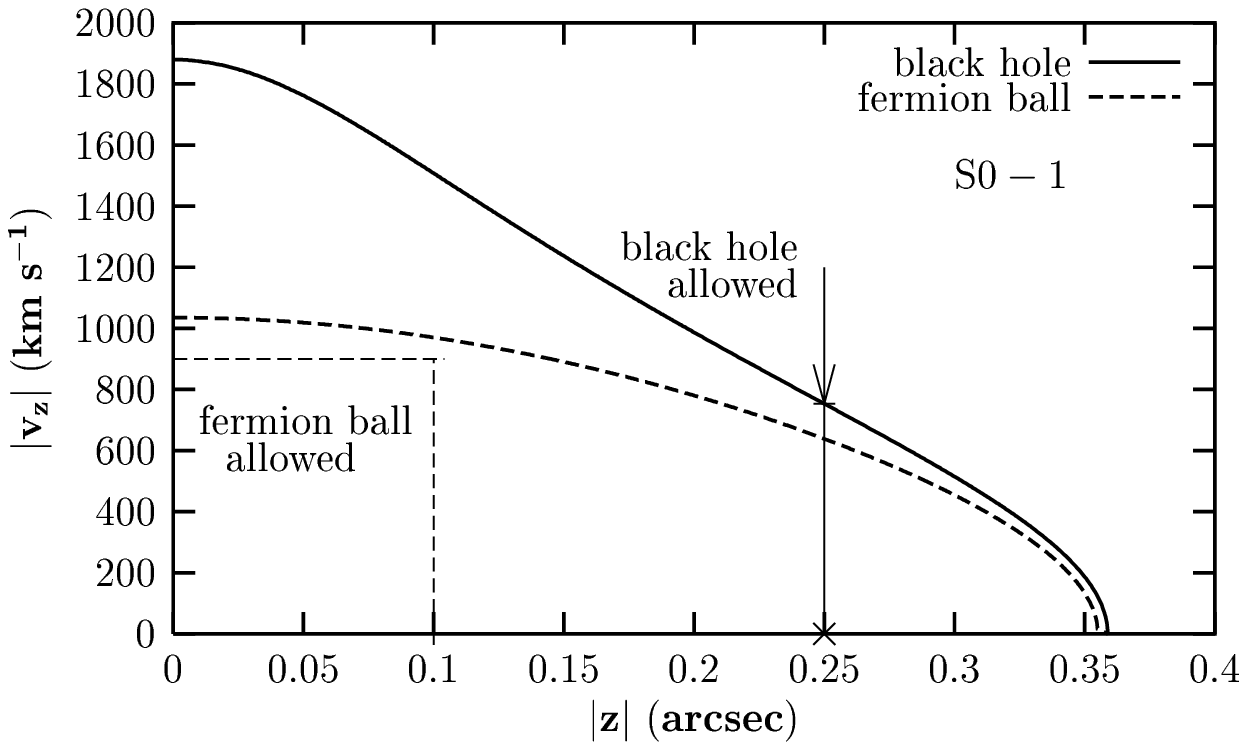}
\end{center}
\caption{
The  $|z|-v_{z}$
phase space
that fits the S0-1 data.
}
\label{fig3a}
\end{figure}
Fig.\ 4 shows some typical projected orbits of S0-1 in the black hole and
fermion ball scenarios. The data of S0-1 may be fitted in both scenarios with
appropriate choices of $v_{x}, v_{y}, z$ and $v_{z}$ in 1995.4. The
inclination angles of the orbit's plane, $\theta$ = arccos
($L_{z}/|\vec{L}|$), with $\vec{L} = m \vec{r} \times \dot{\vec{r}}$, are shown
next to the orbits. The minimal inclination angle that describes the data in
the black hole
case is $\theta$ = 70$^{o}$, while in the fermion ball scenario it is $\theta$ =
0$^{o}$. In the black hole case, the minimal and maximal distances from Sgr
A$^{*}$ are $r_{\rm min}$ = 0.25'' and $r_{\rm max}$ = 0.77'', respectively,
for the orbit with $z$ = 0.25'' and $v_{z}$ = 0 which has a period of $T_{0}
\approx$ 161 yr. The orbits with $z$ = 0.25'' and $v_{z}$ = 400 km s$^{-1}$ or
$z$ = 0.25'' and $v_{z}$ = 700 km s$^{-1}$ have periods $T_{0} \approx$ 268 yr
or $T_{0} \approx$ 3291 yr, respectively. In the fermion ball scenario, the
open orbit with $z$ = 0.1'' and $v_{z}$ = 0 has a ``period'' of $T_{0}
\approx$ 77 yr with $r_{\rm min}$ = 0.13'' and $r_{\rm max}$ = 0.56''.
The open orbits with $z$ = 0.1'' and $v_{z}$ = 400 km s$^{-1}$ or $z$ = 0.1''
and $v_{z}$ = 900 km s$^{-1}$ have ``periods'' of $T_{0} \approx$ 100 yr or
$T_{0} \approx$ 1436 yr, respectively.
\begin{figure}[h]
\begin{center}
\includegraphics[width=.8\textwidth]{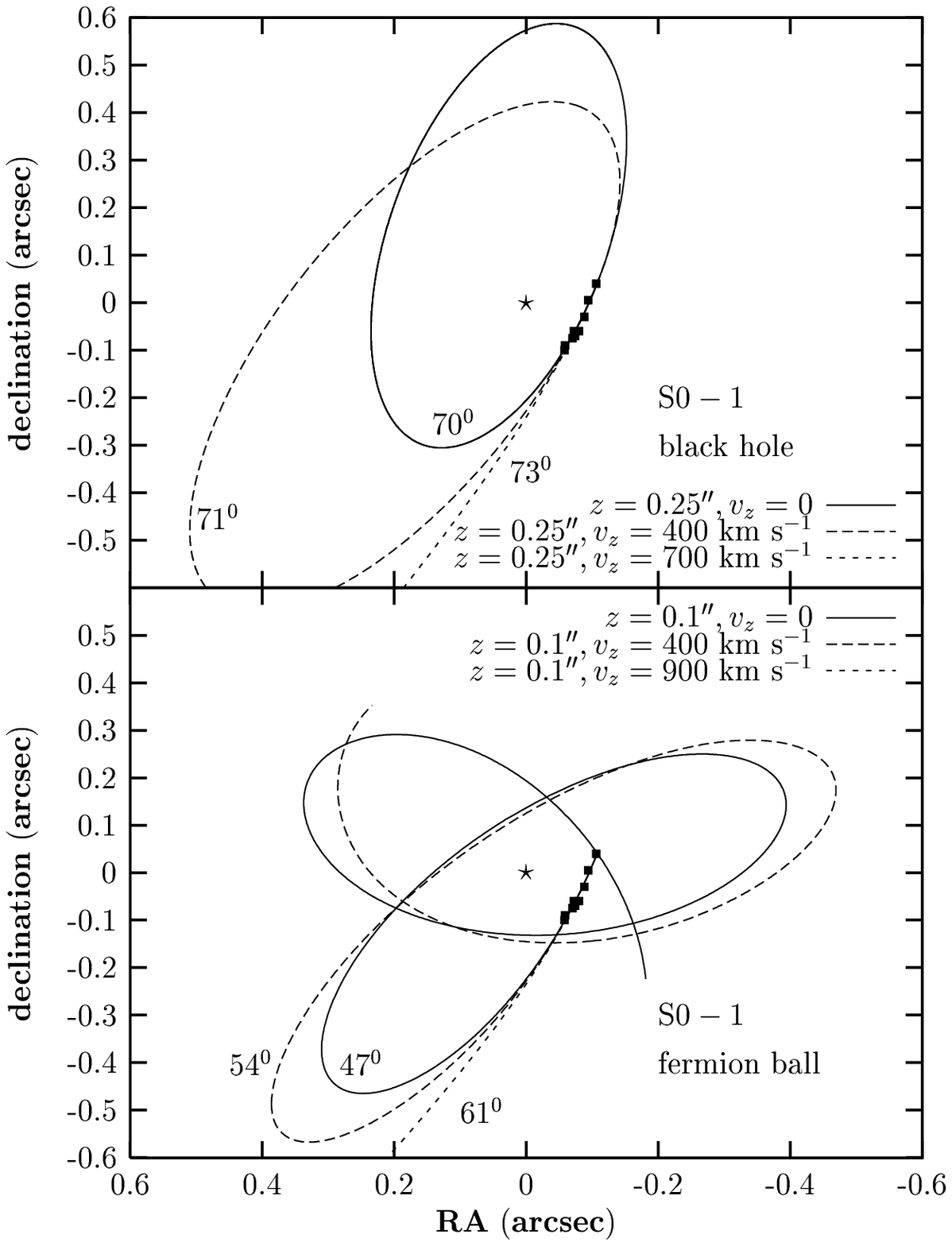}
\end{center}
\caption{
Examples of typical orbits
of S0-1.
}
\label{fig4a}
\end{figure}
\section{Implications and Speculations}
A fermion ball at the Galactic
center could be indirectly observed through the radiative decay of the fermion
(assumed here to be a sterile neutrino) into a standard neutrino, i.e. $f
\rightarrow \nu \gamma$.
If the lifetime for the decay $f \rightarrow \nu \gamma$
is 2.6 $\times$ 10$^{19}$ yr, the luminosity of a $M_{\rm c}$ = 2.6 $\times$
10$^{6} M_{\odot}$ fermion ball would be 2.8 $\times$ 10$^{33}$ erg s$^{-1}$.
This is consistent with the upper limit of the X-ray luminosity
for the quiescent-state
$\sim$ 2.8 $\times$ 10$^{33}$ erg s$^{-1}$ of the source with radius 0.5'' $\simeq$
23 light-days, whose center nearly coincides with Sgr A$^{*}$, as seen by the
Chandra satellite in the 2 to 7 keV band \cite{bag15}. The lifetime is
proportional to sin$^{-2}$ $\theta$, $\theta$ being the unknown mixing angle of
the sterile with active neutrinos. With a lifetime of 2.6 $\times$ 10$^{19}$
yr we obtain an acceptable value for the mixing angle squared of $\theta^{2}$
= 0.44 $\times$ 10$^{-11}$. The X-rays originating from such a radiative decay
would contribute at least two orders of magnitude less than the observed diffuse
X-ray background luminosity at this wavelength if the sterile neutrino is the dark matter
particle of the Universe. The signal observed at the Galactic center would be
a sharp X-ray line at $\sim$ 7.5 keV for $g$ = 2 and $\sim$ 6.3 keV for $g$ = 4. This line could thus be misinterpreted
as the Fe $K_{\alpha}$ line at 6.67 keV.
The X-ray luminosity would be
tracing the fermion matter distribution, and it could thus be an important
test of the fermion ball scenario. Of course the angular resolution would
need to be $\simlt$ 0.1'' and the sensitivity would have to extend beyond 7
keV.

In the fermion ball scenario, the $\sim$ 10 ks X-ray burst observed on 26 October 2000 near Sgr A$^{*}$ would
have to be explained as a thermonuclear instability or runaway of material
accreted or accreting on a neutron star near the center of the fermion ball.
With a neutron star accretion rate of $\sim$ 10$^{-9} M_{\odot}$/yr,
this could also account for
the strong radio emission of Sgr A$^{*}$ in terms of synchroton radiation
due
to $\sim$ 50 MeV
electrons and positrons, produced in $\pi - \mu$ decays,
after inelastic $N + N \rightarrow N + N + \pi$ collisions of the infalling
$\sim$ 200 MeV nucleons hitting the surface of the neutron star. As accreting
matter can easily spin up or slow down the rotation of neutron stars, it is perhaps
also able to keep the neutron star stationary at the center of the fermion
ball for an extended period of time.

In summary, it is important to note that, based on the data of the star
S0-1 \cite{eck10} alone, the fermion ball scenario cannot be ruled out.
Similar results are obtained analyzing the S0-2 data \cite{mun6}.
In fact, in
view of the $z - v_{z}$ phase-space, that is much larger in the fermion ball
scenario than in the black hole case for both the S0-1 and S0-2 data, there is a reason to treat the fermion
ball scenario of the supermassive compact dark object at the center of our
Galaxy with the respect it deserves.
%\subsection*{Acknowledgements}
%This research is in part supported by the Foundation of Fundamental Research
%(FFR) grant number PHY99-01241 and the Research Committee of the University of
%Cape Town. The work of N.B. is supported in part by the Ministry of Science
%and Technology of the Republic of Croatia under Contract No. 0098002.

\end{document}